\documentclass[twocolumn]{aastex63}
\usepackage{amsmath}
\usepackage{color}
\usepackage{ulem}
\usepackage{graphicx}
\usepackage{textcomp}
\usepackage[T1]{fontenc}
\usepackage{footnote}

\begin{document}

\title{Nucleosynthesis Contribution of Neutrino-dominated Accretion Flows to the Chemical Evolution of Active Galactic Nuclei}

\author[0000-0002-1768-0773]{Yan-Qing Qi}
\affiliation{Department of Astronomy, Xiamen University, Xiamen, Fujian 361005, China; tongliu@xmu.edu.cn, msun88@xmu.edu.cn}
\author[0000-0001-8678-6291]{Tong Liu}
\affiliation{Department of Astronomy, Xiamen University, Xiamen, Fujian 361005, China; tongliu@xmu.edu.cn, msun88@xmu.edu.cn}
\author[0000-0002-4223-2198]{Zhen-Yi Cai}
\affiliation{CAS Key Laboratory for Research in Galaxies and Cosmology, Department of Astronomy, University of Science and Technology of China, Hefei, Anhui 230026, China; zcai@ustc.edu.cn}
\affiliation{School of Astronomy and Space Science, University of Science and Technology of China, Hefei, Anhui 230026, China}
\author[0000-0002-0771-2153]{Mouyuan Sun}
\affiliation{Department of Astronomy, Xiamen University, Xiamen, Fujian 361005, China; tongliu@xmu.edu.cn, msun88@xmu.edu.cn}

\begin{abstract}
Recent observations of quasars show high line-flux ratios in their broad emission lines and the ratios appear to be independent of redshift up to $z \gtrsim 6$, which indicate that the broad-line regions of these early quasars are surprisingly metal-rich. Here, we revisit the chemical evolution of high-redshift quasars by adding a new ingredient, i.e., the neutrino-dominated accretion flows (NDAFs) with outflows, on top of the conventional core-collapse supernovae (CCSNe). In the presence of the chemical contribution from NDAFs with outflows, the total metal mass (i.e., the summation of the conventional CCSN and NDAFs with outflows) per CCSN depends weakly upon the mass of the progenitor star if the mass is in the range of $\sim 25-55~M_{\odot}$. We model the chemical evolution by adopting a improved open-box model with three typical initial mass functions (IMFs). We find that, with the additional chemical contribution from NDAFs with outflows, the quasar metallicity can be enriched more rapidly in the very early Universe ($z \sim 10$) and reaches higher saturation than the no-NDAF case at $z \sim 8$, after which they evolve slowly with redshift. The quasar metallicity can reach $\sim 20~Z_{\odot}$ ($Z_\odot$ denotes the metallicity of the Sun; and $\sim 20\%$ of which is produced by NDAF outflows) at $z \sim 8$ for the ``top-heavy'' IMF model in \citet{Toyouchi2022}, which readily explains the quasar observations on the super-solar metal abundance and redshift-independent evolution.
\end{abstract}


\section{Introduction}

High-redshift luminous active galactic nuclei (AGNs), i.e., quasars, provide valuable probes to the metallicity of the early universe, thanks to their high luminosities and various metallic emission lines in the rest-frame ultraviolet (UV) spectra. According to photoionization models \citep[e.g.,][]{Hamann2002,Nagao2006}, UV emission-line ratios like $(\rm Si\,{\textsc {iv}} +O\,{\textsc {iv\scriptsize]}})/C\,{\textsc {iv}}$, $\rm (C\,{\textsc {iii\scriptsize ]}}+Si\,{\textsc {iii\scriptsize]}})/C\,{\textsc {iv}}$, $\rm Al\,{\textsc {iii}}/C\,{\textsc {iv}}$, $\rm N\,{\textsc v}/C\,{\textsc {iv}}$, $\rm He\,{\textsc {ii}}/C\,{\textsc {iv}}$, $\rm Fe\,{\textsc {ii}}/Mg\,{\textsc {ii}}$, and $\rm N\,{\textsc v}/He\,{\textsc {ii}}$ can be used to characterize the broad-line regions (BLRs) metal abundances of high-redshift ($z \gtrsim 6 $) quasars; UV spectroscopic observations identify two noteworthy properties in the BLRs of high-redshift quasars, i.e., super solar metal abundances and no-redshift evolution \citep[e.g.,][]{Jiang2007,DeRosa2014,Wang2022}. Hence, there is a rapid star formation and elemental enrichment within the first gigayear. The supermassive black holes (SMBHs) in these early quasars are already massive ($\sim 10^8-10^{10}~M_{\odot}$) according to near-infrared (NIR) spectroscopic observations \citep[e.g.,][]{Jiang2007,Kurk2007,Wu2015,Yang2021}, indicating a rapid mass growth history.

The gas powering bright AGNs is subject to instability and fragmentation due to self-gravity and a fraction of it might collapse to form local stars, which undergo gas accretion to be massive \citep[e.g.,][]{Goodman2004,Cantiello2021,Jermyn2021}. These stars will explode as supernovae (SNe), polluting the environment with heavy elements and providing a possible explanation for the observed high metal environment in AGNs, and leave behind some stellar remnants. SNe Ia are thermonuclear explosions that originate from a white dwarf (WD) accreting companion or double-WD mergers. They are thought to be one of the most significant nucleosynthetic factories for heavy metals, especially iron-peak elements \citep[see, e.g.,][]{Arnett1996,Hillebrandt2000,McKernan2020}. Massive stars ($\gtrsim 8~M_\odot$) end up as core-collapse SNe (CCSNe) or hypernovae. They are considered to produce the majority of $\alpha-$elements (e.g., oxygen, sulfur, neon, and argon). The nucleosynthesis processes in such energetic SNe has received a great deal of attention \citep[e.g.,][]{Nomoto2006,Heger2010,Sukhbold2016}. Because SNe Ia has a longer evolutionary timescale than CCSNe, iron enrichment from $\alpha-$element is thought to be delayed by $\sim 1~\rm Gyr$. The $\rm Fe\,{\textsc {ii}}/Mg\,{\textsc {ii}}$ flux ratio has been used as a proxy for the $\rm Fe/\alpha-$element abundance. Surprisingly, no significant redshift evolution was discovered up to $z\sim7$, despite the fact that the measurement uncertainty is non-negligible in many cases \citep[e.g.,][]{Kurk2007,DeRosa2014,Wang2022}.

The observed BLR gas is likely to originate from the thermal diffused hot gas continuously supplied by the AGN disk \citep[e.g.,][]{Wang2011}. Shocks from SN explosions interacting with interstellar clouds can lead to the evaporation of clumps and thus the production of hot gas, which is photoionized by the ionizing continua from the central engine and undergoes dynamical and thermal instabilities to drive the formation of the BLR \citep[e.g.,][]{Wang2012}. Therefore, the metal abundance in the AGN disk may determine that in the BLR.

A stellar-mass black hole (BH) with an accretion disk often forms in the center of a collapsar ($\gtrsim 25~M_\odot$). The fall-back matter of the collapsar activates BH hyperaccretion processes, which power relativistic jets through the neutrino-antineutrino annihilation mechanism in the initial accretion phase. The disk that releases the gravitational energy of the BH is named the neutrino-dominated accretion flow \citep[NDAF; for a recent review, see e.g.,][]{Liu2017}. Following the decreasing accretion rates, the Blandford-Znajek mechanism \citep{Blandford1977} that extracts the rotational energy of the BH will dominate over Poynting flux. For NDAFs in the centers of collapsars, the abundance of free protons and neutrons in the disk outflows will participate nucleosynthesis, or even induce the CCSN explosions \citep[e.g.,][]{Pruet2004,Surman2006,Song2019}, and inject the heavy elements to the circumstances \citep{Liu2021}. Such a chemical enrichment process is often disregarded in previous works.

In this paper, we investigate the contribution of CCSNe with or without NDAF outflows to the metallicity evolution of quasars using three typical initial mass functions (IMFs). The paper is organized as follows. In Section 2, we give the metallicity evolution model by considering the gas accretion to the central SMBH, star formation, and SNe (with or without NDAFs) of a quasar. In Section 3, we show the evolution of metallicity within a quasar as a function of IMFs. Conclusions and discussion are made in Section 4. We adopt the cosmology of $\Omega_{0} = 1 - \Omega_{\Lambda} = 0.3$ and $H_{0} = 70~\rm km~ s^{-1}~Mpc^{-1}$ here.

\section{Model}
\subsection{Stellar Yields}

The amount of newly formed and pre-existing elements ejected at the death of all masses of stars, known as stellar yields, is a key parameter in calculating galactic chemical evolution. Numerous studies find that the production and ejection of stellar elements are a function of the initial mass and chemical composition of the star \citep[e.g.,][]{Heger2010,Nomoto2013}. Furthermore, if massive stars ($\gtrsim 25~M_\odot$) explode as CCSNe, the central core falls back to the remnant BHs with accretion disks. The collapsar initial mass-supply rates can keep the accretion processes in the NDAF phase. The metallicity of the progenitor stars determines the electron fraction $Y_{\rm e}$ at the outer boundary of NDAFs \citep[e.g.,][]{Surman2005,Surman2006,Liu2021}. \citet{Liu2012} and \citet{Liu2014} discovered that the Bernoulli parameter is positive especially in the middle and outer regions of the NDAF, implying that outflows may occur, which is similar to the classical advection-dominated accretion flows \citep[e.g.,][]{Gu2009}. Furthermore, many numerical simulations find strong NDAF outflows \citep[e.g.,][]{Fernandez2015,Just2015,Siegel2017}. The initial density, temperature, and materials available for synthesis, all of which are dependent on the state of the disk, are the main parameters that influence the nucleosynthesis in disk outflows. CCSNe and NDAF outflows have the very similar electron fractions (i.e., relative proton-rich condition). Hence, the nucleosynthesis in NDAF outflows is expected to be similar to that in CCSNe \citep[e.g.,][]{Surman2011,Liu2021}. \citet{Liu2021} found that the density profiles and temperatures of the NDAF with outflows are high enough to trigger and maintain the nucleosynthesis process. And they calculated the $^{56}\rm Ni$ yields supplied by CCSNe and NDAF outflows and found that the total $^{56}\rm Ni$ yields are independent of the mass of the progenitor ($\sim 25-55~M_{\odot}$). Similarly, we assume that the total metal yields of CCSNe with NDAFs as a constant value for a progenitor star in that mass range. The stellar yields for the case of no NDAF outflows are taken from the Online Yields Table (2013)\footnote{$\rm http://star.herts.ac.uk/~chiaki/works/YIELD\_CK13.DAT$}.

\begin{figure}
\centering
\includegraphics[width=1.0\linewidth]{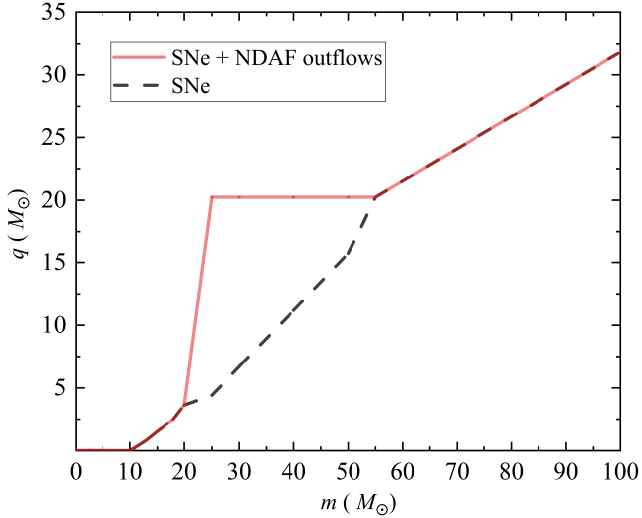}
\caption{Total metal yields from SNe with (red solid line) or without (black dashed line) contribution of NDAF outflows as a function of progenitor star mass.}
\label{totai metal and IMF}
\end{figure}

Figure~\ref{totai metal and IMF} shows the total metal yields $q(m)$ ejected by SNe with (the red solid line) or without (the black dashed line) the NDAF outflows as a function of the progenitor mass for $Z=0$ \citep{Nomoto2013}. The metal loading processes caused by stellar winds from giant stars or compact binary mergers are not taken into account.

\subsection{IMFs\label{IMFs}}

The total yields from a generation of stars depend on the IMF. Although the formation of the first stars has been studied theoretically, their IMFs have yet to be clarified. Moreover, little is known about how the IMF changes as a function of galactic conditions. Here, we make an approximation, that is the IMF is time-independent, i.e., $\phi(m,t)=\phi(m)$, where $m$ denotes the stellar mass in units of the solar mass. The most common parameterizations of the IMF are either a single power law or a power law with multiple breaks, which is defined by $\phi(m)\equiv dN/dm\propto m^{-\alpha_{i}}$ in a mass range of $0.1-100~M_{\odot}$, where $\alpha_{i}$ is the index of the IMF and $0.1~M_{\odot}$ and $100~M_{\odot}$ are the lower and upper limits of the stellar masses, respectively, and it is normalized as
\begin{equation}
\int^{100}_{0.1} m\phi (m)dm=1.
\end{equation}

We quote three typical IMF types here, i.e., \citet{Toyouchi2022} characterized a top-heavy IMF with a single slope $\alpha=-0.5$ (hereafter the Toyouchi IMF), which describe the stellar population in AGN disk; \citet{Zhang2018} assumed a bottom-heavy IMF with a single slope $\alpha=2.7$ (hereafter the Zhang IMF); and \citet{Ballero2007} approximated the IMF with $\alpha_{1}=1.3$ for $0.1\leq m \leq0.5$, and $\alpha_{2}=1.95$ for $0.5\leq m \leq100$ (hereafter the Ballero IMF). In the above IMFs, $\sim 25-55~M_{\odot}$ stars have the highest percentage in Toyouchi IMF, and, the lowest percentage in Zhang IMF.

\subsection{The Metallicity Equation}

Some galaxy chemical evolution models \citep[e.g.,][]{Tinsley1980}, which usually consider gas flows and dilute nucleosynthesis products with unenriched material from outside the galaxy, aim to explain the composition of galaxies based on the stellar yields and the mixing of stellar ejecta with interstellar gas. We improve this chemical evolution model and apply it to the the chemical evolution of the gas of a high-redshift quasar, named the ``open-box'' chemical evolution model, which uses SNe and NDAF outflows as chemical enrichment sources and has only gas accretion to the central SMBH without gas inflow from the environment. Therefore, this can emphasize the role of nucleosynthesis in the NDAF outflows as an additional metal source and reduce the dependence on the composition of outside.

The quasars are powered by the process of accreting materials onto the central SMBH and emit near the Eddington limit. The mass accretion rate is parameterized in terms of the Eddington accretion rate $\dot M_{\rm Edd}=L_{\rm Edd}/\eta c^{2}$, where the Eddington luminosity is provided by $L_{\rm Edd} = 4\pi GM_{\rm smbh}m_{\rm p}c/\sigma_{\rm T}$, $M_{\rm smbh}$ is the SMBH mass, $\sigma_{\rm T}$ is the Thompson cross section for electron scattering, and $m_{\rm p}$ is the proton mass. Here, we take the assumed radiative efficiency $\eta= 0.1$, and thus get SMBH accretion rate at $\lambda$ times Eddington rate to grow the central SMBH,
\begin{equation}
\dot M_{\rm smbh} \approx \frac{dM_{\rm smbh}}{dt}= \lambda \dot M_{\rm Edd}.
\end{equation}

Then, the equations describe the evolution of star formation, gas, and metal abundance as follows
\begin{equation}
\dot \Psi = M_{\rm g} ~ / ~ t_\star,
\end{equation}
\begin{equation}
\frac{d M_{\rm g}}{dt} = - \dot \Psi + R - \dot M_{\rm smbh},
\end{equation}
\begin{equation}
\frac{d(M_{\rm g} X)}{dt} = - \dot \Psi X + E - \dot M_{\rm smbh} X,
\end{equation}
where $\dot \Psi$ is the star formation rate \citep[SFR; e.g., see Equation~(A4) of][]{Cai2013}, $M_{\rm g}$ is the total mass of gas, $t_\star$ is a typical star formation timescale, $R$ is the rate of mass ejection at the end of stellar evolution, $X$ is the mass fraction of metals, and $E$ is the rate of metal ejection from SNe with or without NDAF outflows, respectively.

We use the instantaneous recycling approximation, i.e., a star was formed at time $t - \tau(m)$ if it dies at time $t$, and $\tau(m)$ is the lifetime of stars with mass $m$. The mass ejection rate at time $t$ is
\begin{equation}
R(t) = \int_{m_{\star}(t)}^{m_{\rm max}} (m - m_{\rm rem}) ~ \phi(m) ~ \dot \Psi[t - \tau(m)] dm,
\end{equation}
where $m_{\rm rem}(m)$ is the remnant mass, and $m_\star$ is the mass of the star for which $\tau(m_\star) = t$. The remnant mass $m_{\rm rem}(m)$ is provided by the Online Yields Table (2013). The lifetime of a star with mass $m$ is estimated according to the work of the Geneva group \citep{Schaller1992}
\begin{equation}
\tau(m) \approx 11.3 m^{-3} + 0.06 m^{-0.75} + 0.0012~~{\rm Gyr}.
\end{equation}

The overall amount of metals produced by SNe with or without NDAFs is
\begin{equation}
\begin{aligned}
E(t)\simeq &\int_{m_{\star}(t)}^{m^{\rm max}}\ q(m)~\phi(m)~\dot \Psi[t - \tau(m)]~dm,
\end{aligned}
\end{equation}
where $q(m)$ is the metal yield as shown in Figure~\ref{totai metal and IMF}. We further assume that the initial metallicity of the star has no effect on $q(m)$, i.e., the metal under consideration is a primary element. Metal production of the progenitor mass with $Z=0$ and $Z_\odot$ is compared \citep[see Figure 4 in][]{Nomoto2013}, which slightly decreases at high metallicity.

\subsection{Observational consequences for the BLR}

We collect 24 quasars at $z \gtrsim 5.7$ from \citet{Jiang2007} and \citet{Wang2022} and utilize the theoretical relation $\textrm{log}~Z/Z_{\odot} =1.0 + 1.33~\log (\rm N\,{\textsc v}/C\,{\textsc {iv}})$ to convert the line-flux ratios $\rm N\,{\textsc v}/C\,{\textsc {iv}}$ to metallicities $Z$ \citep[][]{Hamann2002,Wang2010}. These results are shown in Figure~\ref{X}, which imply that quasar BLRs were enriched in metals at $z\sim6$, and that the enrichment of BLR metal abundances must have happened much earlier and/or much faster. The conversion of line ratios to metallicities has substantial systematic uncertainties. The ionizing parameter, the hardness of the ionizing continuum, temperature, density, and other parameters can all have effects on the conversion \citep[see][and reference therein]{Maiolino2019}. Considering only show a comparison of the present model with the observed samples in high-redshift quasars, we will not attempt to understand the differences between each metal abundance indicator in this work, nor will we reach any conclusions about which should be used as the true metal abundance.

\section{Metallicity within the quasar}

\begin{figure}
\centering
\includegraphics[width=1.0\linewidth]{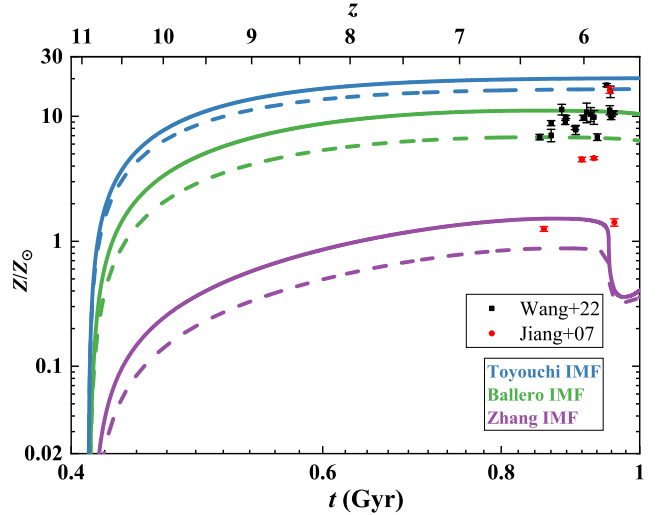}
\caption{Metallicity evolutions in scenarios of SNe with (solid lines) or without (dashed lines) NDAF outflows by using yields and three typical IMFs described in Section~\ref{IMFs}. Different observed samples are from Wang+22 \citep[black rectangles;][]{Wang2022} and Jiang+07 \citep[red circles;][]{Jiang2007}.}
\label{X}
\end{figure}

The primary mechanisms for the seed formation of SMBHs are the collapse of a massive star, runaway collisions in stellar clusters, and the direct collapse of a gas cloud into a massive BH \citep[for a review, see][]{Latif2016}. Based on the measured bolometric luminosities and BH masses, \citet{Yang2021} derived the Eddington ratios of high-redshift quasars with a mean of 1.08. We begin by assuming that the seed BH with $100~M_{\odot}$ at $z\sim 11$ grows in the Eddington limit $\lambda=1$ until the accretion rate changes to zero when the gas surrounding the SMBH is about to be consumed. For a quasar with initial gas mass $M_{\rm g,0}=10^{9}~M_{\odot}$ \citep[e.g.,][]{Storchi-Bergmann2019}, and star formation timescale $t_\star= 0.1~\rm Gyr$, we show the chemical evolution altered with different IMF types in Figure~\ref{X} by considering the SNe with or without nucleosynthesis of NDAF outflows.

As shown in Figure \ref{X}, if the NDAF outflow nucleosynthesis is added, the metal abundance can be rapidly raised to higher values at $z\sim 10$ than the no-NDAF case. This is because in the very early Universe, only massive and short-lived stars end their lives and the accompanying NDAF outflow nucleosynthetic is prominent. The metallicity increases with time until it reaches a saturation value at $z \sim 8$. If the SNe and NDAF outflow contributions are considered simultaneously, the metal abundances can grow to $20.2$, $10.6$, and $1.0~Z_{\odot}$ for the IMFs adopted from Toyouchi IMF (blue solid line), Ballero IMF (green solid line) and Zhang IMF (purple solid line), respectively. However, if only SNe explosions are considered, the gas can be enriched to $\sim16.5~Z_{\odot}$ in Toyouchi IMF (blue dashed line), $\sim6.5~Z_{\odot}$ in Ballero IMF (green dashed line) but only to $\sim0.7~Z_{\odot}$ in Zhang IMF (purple dashed line). NDAF outflows act as the important sources of nucleosynthesis, resulting in a higher metal abundance ratio in a short period of time and a higher (by a factor of $\sim 120\%$) saturation value, which naturally explains all the observational data for the Toyouchi IMF case, including those the super solar metal abundance and non-redshift evolution of high-redshift quasars.

In addition, the metallicity is affected by the choice of popular IMFs, which typically offer various ratios of the massive stars. It is worth noting that the metallicity is greatest when the IMF is Toyouchi IMF \citep[blue lines;][]{Toyouchi2022} and lowest when the IMF is the Zhang IMF \citep[purple lines;][]{Zhang2018}. This is because stars with the mass of $25-100~M_{\odot}$ have the highest ratio in the IMF for Toyouchi IMF and the lowest ratio in the IMF for Zhang IMF. For the Toyouchi IMF, the metallicity is already high enough to explain all the observations even without taking into account the contribution of NDAF outflows. The Toyouchi IMF supports the frequent occurrence of massive stars, so that the stellar mass distribution is expected to be ``top-heavy'' \citep[e.g.,][]{Goodman2004,Mapelli2012}. Such a special IMF probably only holds for stars in quasar disks \citep[e.g.,][]{Toyouchi2022}. Observations of the central nuclear region ($\sim0.5\rm ~pc$) of the Milky Way reveal the presence of hundreds of massive young OB stars and Wolf-Rayet stars \citep[e.g.,][]{Levin2003}, with no low-mass stars, which is contrary to the number predicted by the standard Salpeter IMF. Compare the predictions of the model with observations of chemical abundance ratios in quasars, constraints can be put on the IMFs.

AGN activity and star formation are heavily reliant on the availability of cold gas, which is the fuel. Because massive stars make up a large proportion in the top-heavier IMF, the death of massive stars can return a large amount of gas in time to keep the mass of quasar system at a high value and maintain the SMBH accretion for a longer time. As \cite{Gao2004} points out, SFR scales linearly with the dense molecular  gas mass. Hence, the top-heavier the IMF, the higher the SFR. With Eddington accretion and radiation efficiency of 0.1 in Toyouchi IMF, it would take $\sim \rm 0.7~Gyr$ for a seed BH of $100~M_{\odot}$ to grow to a $\sim 10^{9}~M_{\odot}$ SMBH, which is more massive than those in the cases of Ballero IMF and Zhang IMF, and within the observed SMBH mass range ($\sim 10^{8}-10^{10}~M_\odot$).

\section{Conclusions and Discussion}

In this paper, we calculate the metallicity evolution of high-redshift quasars with different IMFs by considering the contributions of SNe with or without NDAF outflows. We find that if the contributions of NDAF outflows are added, the metal abundance can be raised to a higher value at an earlier time than the no-NDAF case, thereby offering in a natural explanation for the unexpectly high abundance of metals in high-redshift quasars. In addition, the metallicity reaches its maximum value at $z\sim 8$, which can be increased by at least $\sim 20\%$ compared to the no-NDAF case, and then evolves slowly with the redshift. During an episode of the SMBH activity, the quasar metallicity can be used as a ``clock'' to estimate its age.

Metal-rich environments are fully established when the IMF is ``top-heavy'', which supports the frequent occurrence of massive stars. For the Toyouchi IMF, the metal abundance can reach more than ten times the solar metal abundance even without accounting for the contribution of NDAF outflows. The nucleosynthesis contributions of NDAF outflows also significantly increase the metal abundance by $\sim 20\%$ and should be reasonably included in the chemical evolution of galaxies and AGNs.

The measurements of the metallicity in quasar BLR may lead to other intriguing findings since the BLR metallicity depends upon either SMBH mass or quasar luminosity \citep[e.g.,][]{Warner2003,Nagao2006,Matsuoka2011,Xu2018}, and thus can shed light on the co-evolution of galaxies and their central SMBHs.

\acknowledgments
This work was supported by the National Natural Science Foundation of China under grants 12173031, 11873045, and 11973002, and the science research grants from the China Manned Space Project with Nos. CMS-CSST-2021-A06 and CMS-CSST-2021-B11.

\end{document}